\documentclass[letter]{jpsj3}
\usepackage{graphicx}

\title{Anomalous Spin Dynamics Observed by High-Frequency ESR in Honeycomb Lattice Antiferromagnet InCu$_{2/3}$V$_{1/3}$O$_3$}

\author{Susumu Okubo$^1$\thanks{E-mail address: sokubo@kobe-u.ac.jp}, Hideo Wada$^2$, Hitoshi Ohta$^1$, 
Takahiro Tomita$^{1}$\thanks{Present address: Department of Physics, College of Humanities and Sciences, Nihon University, Sakurajosui, Setagaya-ku, Tokyo 156-8550, Japan}, 
Masashi Fujisawa$^{1}$\thanks{Present address: Department of Applied Physics, University of Fukui, 3-9-1 Bunkyo, Fukui 910-8507, Japan}, 
Takahiro Sakurai$^3$, Eiji Ohmichi$^2$, Hikomitsu Kikuchi$^4$}
\inst{$^1$Molecular Photoscience Research Center, Kobe University, 1-1 Rokkodai-cho, Nada, Kobe 657-8501, Japan\\
$^2$Graduate School for Science, Kobe University, 1-1 Rokkodai-cho, Nada, Kobe 657-8501, Japan\\
$^3$Center for Support to Research and Education Activities,  Kobe University, 1-1 Rokkodai-cho, Nada, Kobe 657-8501, Japan\\
$^4$Department of Applied Physics, University of Fukui, 3-9-1 Bunkyo, Fukui 910-8507, Japan}

\abst{High-frequency ESR results on the $S$ = 1/2 Heisenberg hexagonal antiferromagnet InCu$_{2/3}$V$_{1/3}$O$_3$ are reported. 
This compound appears to be a rare model substance for the honeycomb lattice antiferromagnet with very weak interlayer couplings. 
The high-temperature magnetic susceptibility can be interpreted by the $S$ = 1/2 honeycomb lattice antiferromagnet, and it shows a magnetic-order-like anomaly at $T_{\rm N}$ = 38 K. 
Although, the resonance field of our high-frequency ESR shows the typical behavior of the antiferromagnetic resonance, the linewidth of our high-frequency ESR continues to increase below $T_{\rm N}$, while it tends to decrease as the temperature in a conventional three-dimensional antiferromagnet decreases. 
In general, a honeycomb lattice antiferromagnet is expected to show a simple antiferromagnetic order similar to that of a square lattice antiferromagnet theoretically because both antiferromagnets are bipartite lattices. 
However, we suggest that the observed anomalous spin dynamics below $T_{\rm N}$ is the peculiar feature of the honeycomb lattice antiferromagnet that is not observed in the square lattice antiferromagnet. }

\kword{antiferromagnet, high field, ESR, AFMR, honeycomb lattice, spin dynamics}

\begin{document}
\maketitle

Dimensionality and the spin magnitude $S$ play important roles in the physical properties of interacting spin systems because they significantly affect quantum fluctuation. 
For instance, one-dimensional (1D) Heisenberg antiferromagnets show the Haldane gap in the case of $S$ = 1 owing to the quantum effect \cite{Haldane}, while the $S$ = 1/2 system does not show the gap except for the spin-Peierls system in the case of strong spin-phonon coupling \cite{Huizinga,Hase}. 
In the study of model substances of such exotic systems, high-frequency, high-field ESR measurements played important roles \cite{Date,Lu,Yoshida,Ohta1,Brill,Nojiri}. In two-dimensional (2D), two simple bipartite lattices with nearest-neighbor antiferromagnetic (AF) interactions, namely, a square lattice and a honeycomb lattice, exist where they undergo AF order in the large $S$ limit \cite{Takano1}. 
On the other hand, the quantum fluctuation for $S$ = 1/2 enhanced by the frustration and distortion of interactions may destroy the AF order. 
However, theoretical studies about the existence of the disordered state in the $J_1$-$J_2$ model of the $S$ = 1/2 square lattice are still controversial \cite{Takano1}. 
In contrast, the AF order in the $S$ = 1/2 honeycomb lattice is more fragile to the quantum fluctuation because its coordination number 3 is smaller than 4 of the square lattice. 
Therefore, there are many theoretical studies of honeycomb lattices \cite{Takano1,Einarsson,Chakravarty,Takano2,Mattsson,Fout}. 
In particular, Takano studied the honeycomb lattice antiferromagnet with frustration by the second-neighbor AF interaction $J_2$ and dimer-like distortion in the nearest-neighbor AF interaction $J_1$ by the nonlinear   model (NLSM) \cite{Takano1}. 
He obtained a ground state phase diagram consisting of an ordered AF phase and a disordered spin-gap phase, and suggested that the spin-gap phase for a honeycomb lattice is larger than that for the $J_1$-$J_2$ model on a square lattice, in the case of $S$ = 1/2 supporting the fragileness of AF order in a honeycomb lattice. \cite{Takano1}. 
Therefore, the spin dynamics near  $T_{\rm N}$ may be different between honeycomb and square lattice antiferromagnets, and the ESR measurement conducted to find the peculiar spin dynamics in the honeycomb lattice antiferromagnet will be a very interesting issue.

\begin{figure}
\begin{center}
\includegraphics[bb=0 0 1570 1082,keepaspectratio=true,width=70mm]{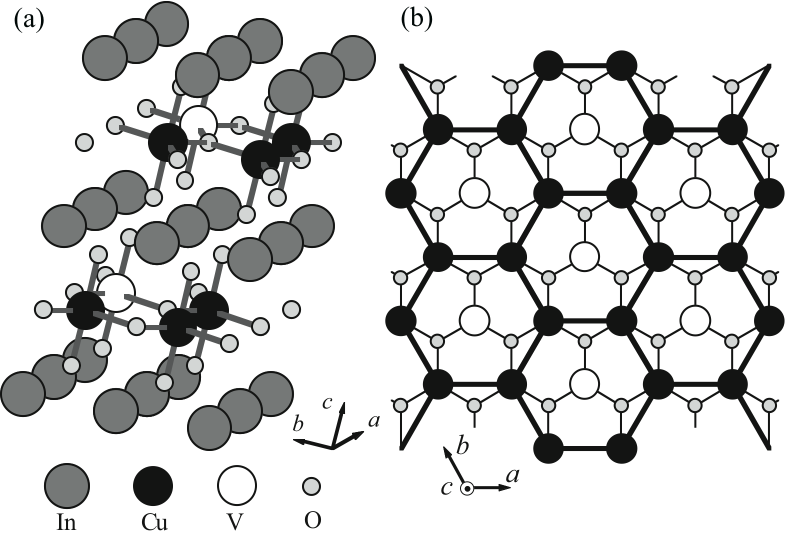}
\end{center}
\caption{(a) Crystal structure of InCu$_{2/3}$V$_{1/3}$O$_3$. (b) Honeycomb Cu$^{2+}$ network in the $c$-plane.}
\label{fig:fig1}
\end{figure}

Although the model substance is important for the experimental study as shown in the case of 1D systems, the model substance is rare for the honeycomb lattice. 
The only known model substances for the honeycomb lattice are InCu$_{2/3}$V$_{1/3}$O$_3$  \cite{Kataev} and Na$_3$T$_2$SbO$_6$ \cite{Miura}, while InCu$_{2/3}$V$_{1/3}$O$_3$ is the only system with equivalent exchange interactions. 
The crystal structure of InCu$_{2/3}$V$_{1/3}$O$_3$ has a hexagonal space group of $P6_3/mmc$ with the lattice parameters $a$ = 0.33564 nm and $c$  = 1.1908 nm \cite{Kataev}. 
The magnetic ion is Cu$^{2+}$ ($S$ = 1/2), which is surrounded by a trigonal-bipyramidal coordination of oxygen ions, as shown in Fig. \ref{fig:fig1}(a). 
The Cu$^{2+}$ network in the $c$-plane is expected to take a honeycomb structure, as shown in Fig. \ref{fig:fig1}(b), because the nonmagnetic V$^{5+}$ ion is expected to be surrounded by hexagons of Cu ions owing to its larger positive charge and smaller size. 
As each honeycomb layer is well separated from other honeycomb layers by nonmagnetic In$^{3+}$ ions along the $c$-axis, the system can be considered as a quasi 2D system. 
The magnetic susceptibility measurement of InCu$_{2/3}$V$_{1/3}$O$_3$ powder at 1 T by Kataev \textit{et al}. \cite{Kataev} showed a broad maximum at 180 K, which is a characteristic of a low-dimensional antiferromagnet, and a kink at 38 K with a Curie-like upturn at a lower temperature. 
From the analysis results, the lower temperature behavior can be interpreted by the Curie-Weiss contribution of about 3\% of paramagnetic $S$ = 1/2 moments with the Weiss temperature of 10 K and the small contribution of the Van-Vleck term. 
Moreover, the overall feature of the magnetic susceptibility can be understood on the basis of the susceptibility of an $S$ = 1/2 Heisenberg antiferromagnet on the honeycomb lattice while the deviation from the experimental data becomes visible below 80 K. 
The calculation was performed by the quantum Monte-Carlo Loop algorithm, and the obtained nearest-neighbor exchange parameter of the Heisenberg Hamiltonian ${\mathcal H} = 2J \sum {\bf S}_i \cdot {\bf S}_j$ was $J$ = 140 K \cite{Kataev}. 
However, the three-dimensional (3D) antiferromagnetic order was suggested at $T_{\rm N}$ = 38 K as a kink in the magnetic susceptibility, and it was also supported by the X-band (9.47 GHz) ESR results, that showed the divergent broadening of the linewidth at $T_{\rm N}$ and the disappearance of ESR below $T_{\rm N}$ \cite{Kataev}.
However, the antiferromagnetic resonances below $T_{\rm N}$ could not be observed in the X-band ESR measurements, because the antiferromagnetic resonances shifted away from the measured magnetic fields as a result of an antiferromagnetic gap that is larger than X-band frequency. 
Moreover, as Kataev {\it et al.} treated the powder sample, their obtained linewidths included extrinsic values from anisotropic g-values. 
In this paper, to obtain the spin dynamics of the S=1/2 honeycomb lattice antiferromagnet InCu$_{2/3}$V$_{1/3}$O$_3$, high-frequency ESR results are presented, and the anomalous spin dynamics observed below $T_{\rm N}$ is reported and discussed as a peculiar feature of the honeycomb lattice antiferromagnet.

High-field ESR measurements of powder and magnetically aligned samples of InCu$_{2/3}$V$_{1/3}$O$_3$ have been performed in the temperature region from 1.8 to 265 K using pulsed magnetic fields up to 16 T. 
Gunn and backward traveling wave oscillators, which cover the frequency region from 60 to 600 GHz, have been used as light sources. 
Details of our high-field ESR systems can be found in refs. 18-21. 
A powder sample of  InCu$_{2/3}$V$_{1/3}$O$_3$ is obtained from mixtures of In$_2$O$_3$, CuO and V$_2$O$_5$ (molar ratio of 3:4:1) at 900 $^{\circ}$C for 5 days, which is similar to the previous procedure \cite{Kataev}. 
The obtained powder sample is examined by powder X-ray measurement. 
As the single crystal InCu$_{2/3}$V$_{1/3}$O$_3$ is not currently available, the magnetically aligned sample is prepared by mixing the powder sample and epoxy resin at a static field of 10 T at room temperature. 
To apply the static field, the He-free magnet  manufactured by JASTEC is used. 
After 5 to 24 hours, the epoxy resin hardens and the magnetically aligned sample is produced.

\begin{figure}
\begin{center}
\includegraphics[bb=0 0 1044 852,keepaspectratio=true,width=60mm]{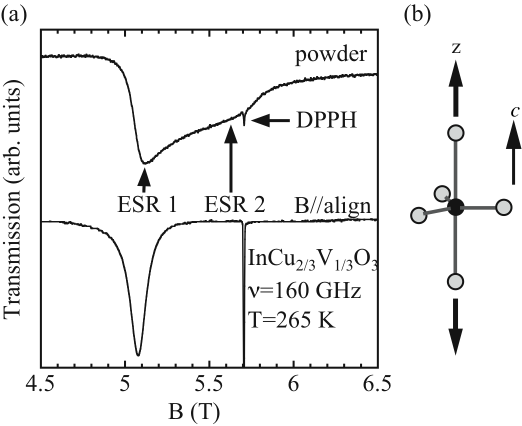}
\end{center}
\caption{(a) ESR spectra of powder and magnetically aligned samples for $B$ // aligned direction observed at 265 K. DPPH is the field marker that corresponds to $g$=2. (b) Schematic figure of principle and crystal axes.}
\label{fig:fig2}
\end{figure}
In the powder-pattern ESR spectrum shown in Fig. \ref{fig:fig2}, where it reflects the integration of ESR over the entire magnetic field angles, the peaks ESR1 and ESR2 correspond to the cases when the magnetic field is applied to the principal axes \cite{Ohta3}. 
The separation of ESR1 and ESR2 is due to anisotropic g-values. 
The separation of ESR1 and ESR2 was not possible in the previous powder X-band ESR experiment, because of the small anisotropy of g-values. 
However, it is possible in the case of using a higher frequency because of the higher spectral resolution of high-frequency ESR \cite{Ohta4}. 
The $g$-values of ESR1 and ESR2 are obtained from the frequency-field measurement at 265 K. 
The obtained $g$-values are $g_1$ = 2.24 $\pm$ 0.01 and $g_2$ = 2.02 $\pm$ 0.01. 
Since the spectral weight of ESR1 is larger than that of ESR2 as shown in Fig. \ref{fig:fig2}(a),  $g_1$ and $g_2$ can be assigned as $g_{\perp}$  (perpendicular to the $c$-axis) and $g_{//}$ (parallel to the $c$-axis), respectively \cite{Abragam}. 
Therefore, the orbital ground state of InCu$_{2/3}$V$_{1/3}$O$_3$  can be considered as $3z^2-r^2$. 
It should be noted that in most Cu$^{2+}$ systems, the ground state is a $x^2-y^2$ state, where $g_{//}$ is larger than $g_{\perp}$  \cite{Abragam}. 
However, the $3z^2-r^2$ ground state is a characteristic of an elongated bipyramidal crystal field  (Fig. \ref{fig:fig2}(b)) and is supported by the optical measurement and molecular orbital calculation \cite{Kataev}. 
The previous X-band ESR of the magnetically aligned sample also suggested the same conclusion with $g_a$ = 2.24 and $g_c$ = 2.00, which were supported by the X-ray diffraction measurement \cite{Kataev}.

\begin{figure}
\begin{center}
\includegraphics[bb=0 0 1112 1050,keepaspectratio=true,width=75mm]{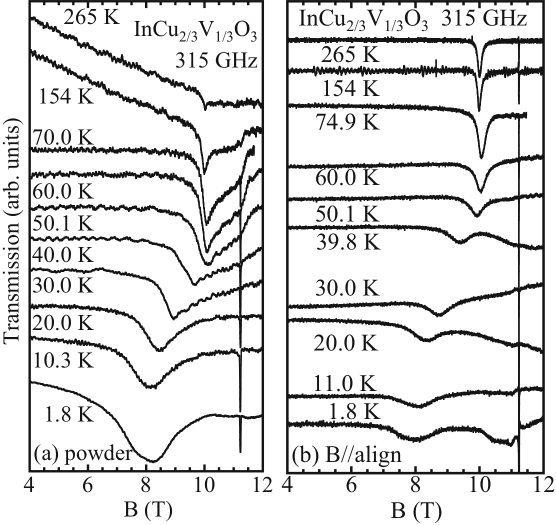}
\end{center}
\caption{ESR spectra of (a) InCu$_{2/3}$V$_{1/3}$O$_3$ powder and (b) $B$ // aligned direction observed at 315 GHz.}
\label{fig:fig3}
\end{figure}

Figure \ref{fig:fig2}(a) shows the comparison of ESR spectra for powder and $B$ // aligned direction at 265 K. 
The result clearly shows that the aligned direction is along $g$ = 2.24, which is consistent with the result obtained by Kataev {\it et al.} \cite{Kataev}. 
One can also expect that the ESR signal for the $B$ // aligned direction will reflect the intrinsic linewidth of InCu$_{2/3}$V$_{1/3}$O$_3$ because Fig. \ref{fig:fig3} shows the well-aligned nature of the magnetically aligned sample.

\begin{figure}
\begin{center}
\includegraphics[bb=0 0 822 810,keepaspectratio=true,width=60mm]{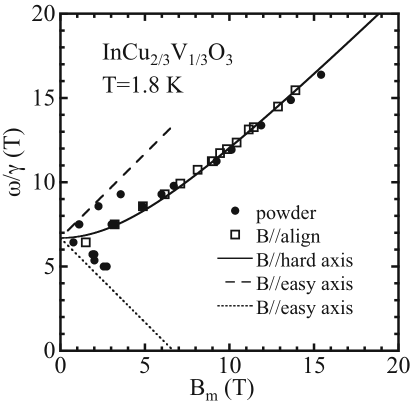}
\end{center}
\caption{Frequency-field diagram of AFMR observed at 1.8 K. Solid symbols and open squares correspond to the observed AFMRs for the powder sample and $B$ // aligned direction, respectively. A solid line ($B$ // hard axis, eq. (1)) and dashed and dotted lines ($B$ // easy axis, eq. (2)) show the AFMR modes expected using the conventional mean field AFMR theory \cite{Ohta3}.}
\label{fig:fig4}
\end{figure}

Figure \ref{fig:fig3} shows the temperature-dependent ESR spectra of powder and $B$ // aligned direction observed at 315 GHz. 
It is clear that the ESR in the paramagnetic region tends to broaden as the temperature approaches $T_{\rm N}$ = 38 K, which is a typical behavior. 
Then, the resonance shifts to a lower field are observed below $T_{\rm N}$, which suggests the development of an internal field. 
Therefore, the resonance observed below $T_{\rm N}$ can be regarded as the anitiferromagnetic resonance (AFMR). 
Here, we should point out that such a measurement below $T_{\rm N}$ was not possible by Kataev {\it et al}. using the X-band ESR \cite{Kataev} owing to the existence of the internal field, and we have observed AFMR for the first time. 
Figure \ref{fig:fig4} shows the frequency-field diagram of the observed AFMR at 1.8 K. 
Although an AF gap of about 180 GHz is clearly observed in the diagram, the identification of AFMR below 140 GHz is difficult owing to the broadness of AFMR. 
As Fig. \ref{fig:fig4} resembles the frequency-field relation of AFMR with a uniaxial anisotropy, the obtained results are analyzed by the following AFMR relations: \cite{Ohta3}
\begin{eqnarray}
&B\ // \ {\rm hard\ axis}, & \ \frac{\omega}{\gamma} = \sqrt{B_{\rm m}^2 + C^2} \\ 
&B\ // \ {\rm easy\ axis}, &\ \frac{\omega}{\gamma} = \pm B_{\rm m} + C
\end{eqnarray}
where $\omega$, $\gamma$, $B_{\rm m}$, $B_{\rm exp}$, and $C$ are the angular frequency, gyromagnetic ratio, modified field, experimental resonance field, and antiferromagnetic (AF) gap, respectively, and $B_{\rm m} = B_{\rm exp} (g/2)$. 
Here, $g$ = 2.24 and $g$ = 2.02 are used for the $B$ // hard and easy axes, respectively. 
The obtained AF gap is $C$ = 180 GHz from the analysis results. 
Moreover, it is clear from Fig. \ref{fig:fig4} that the observed resonances at 1.8 K can be well explained using the AFMR theory, and Figs. \ref{fig:fig3} and \ref{fig:fig4} support the fact that $T_{\rm N}$ = 38 K is the N\'{e}el order. 
However, the spin-flop transition at around 7 T, which can be estimated from the AF gap using the conventional mean field theory \cite{Ohta3}, was not observed by our magnetization measurement at 1.8 K using the pulsed magnetic field. 

\begin{figure}
\begin{center}
\includegraphics[bb=0 0 864 936,keepaspectratio=true,width=60mm]{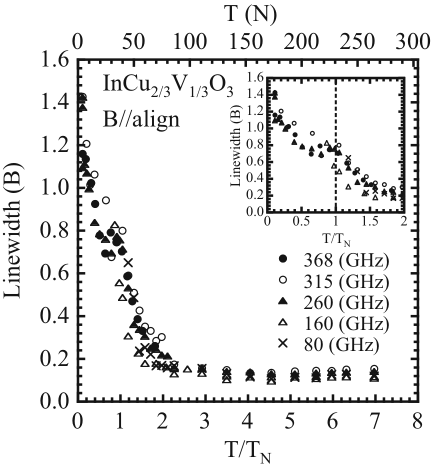}
\end{center}
\caption{Temperature dependence of linewidth in the case of $B$ // aligned direction. Inset shows close up plot at around $T_{\rm{N}}$. }
\label{fig:fig5}
\end{figure}

Although the resonance fields at 1.8 K can be interpreted using the AFMR theory, the linewidth of AFMR seems to be broader as the temperature decreases below $T_{\rm N}$ as shown in Fig. \ref{fig:fig3}, while it becomes sharper as the temperature decreases below $T_{\rm N}$ in the case of the conventional AFMR \cite{Ohta3}. 
Therefore, in order to obtain a better insight into the spin correlation of the system, we have investigated the temperature and frequency dependences of the high-frequency ESR of the magnetically aligned sample in detail. 
Figure \ref{fig:fig5} shows the temperature dependence of ESR linewidth for the $B$ // aligned direction. 
Above $T_{\rm N}$, the linewidth tends to diverge at all frequencies as the temperature approaches $T_{\rm N}$ owing to the spin fluctuation, which is a typical behavior of the antiferromagnet, as discussed by Mori and Kawasaki \cite{Mori}. 
Although such a divergence tends to occur below $T/T_{\rm N}$ = 1.1 from the theory, the linewidth starts to increase below $T/T_{\rm N}$ = 2.5 in our case, which can be understood by the short-range order originating from the low-dimensional nature of our system. 
The level of temperature dependence of AFMR linewidth below $T_{\rm N}$ was discussed by Johnson and Nethercot, \cite{Johnson} and it is expected to decrease as the temperature decreases for the typical antiferromagnet because the spin fluctuation is dominant only near $T_{\rm N}$. 
However, our results in Fig. \ref{fig:fig5} continue to increase as the temperature decreases, which is observed for the first time. 
Such an anomalous behavior below $T_{\rm N}$ is also observed in NMR, where $1/T_1$ does not show a $T^3$ decrease \cite{Fujii}. 
A change in slope is also seen in Fig. \ref{fig:fig5} at 20 K, which may correlate with the small anomaly observed at around 20 K in the $1/T_1$ of NMR \cite{Fujii}. 
Therefore, these results clearly suggest that a certain type of quantum fluctuation still remains below $T_{\rm N}$ in InCu$_{2/3}$V$_{1/3}$O$_3$.

According to the Monte Carlo calculations of $S$=1/2 Heisenberg honeycomb lattice and square lattice antiferromagnets with only the nearest-neighbor exchange interaction $J_1$, the ground-state staggered magnetization of the honeycomb lattice has a much smaller value, $m^\dagger=0.22$ (44\% of the saturation value)\cite{honeycomb_Reger}, than that of the square lattice, $m^\dagger=0.30$ (60\% of the saturation value)\cite{honeycomb_Reger, square_Reger, square_Gross}, which is consistent with the expected fragileness of AF order in the honeycomb lattice originating from the smallest coordination number $z=3$ in the 2D lattice. 
As these theoretical results suggest  the existence of a stronger fluctuation in the $S=1/2$ honeycomb lattice antiferromagnet, the unusual spin dynamics observed in Fig. 5 can be expected. 
Moreover, it is very important to compare our ESR result with that of the $S$=1/2 square lattice antiferromagnet. 
La$_2$CuO$_4$ is a well-known model substance of the $S$=1/2 square lattice antiferromagnet, but its ESR spin dynamics below $T_{\rm N}$ is not known because its AFMR gap is beyond several THz owing to its strong exchange interactions \cite{La2CuO4}. 
However, Ba$_2$Cu$_3$O$_4$Cl$_2$ can be another candidate of the $S$=1/2 square lattice antiferromagnet  \cite{Sreedhar}. 
It has two ${\rm Cu^{2+}}$ sites, which form square lattices in the 2D $c$-plane. 
Noro $et$ $al.$ suggested that ${\rm Cu_B}$ sites, which are coupled by strong 180 degrees Cu-O-Cu superexchange interactions, show a long-range order at $T_{\rm H}$=320 K, but ${\rm Cu_A}$ sites remain paramagnetic even below  $T_{\rm H}$ because the exchange fields from nearest-neighbor ${\rm Cu_B}$ sites are almost canceled out from the structural consideration \cite{Noro}.
Therefore, ${\rm Cu_A}$ sites, which have weaker superexchange interactions, show an antiferromagnetic order below $T_{\rm L}$=40 $K$ \cite{Noro}. 
The submillimeter wave AFMR of the Ba$_2$Cu$_3$O$_4$Cl$_2$ single crystal is observed at 370.4 GHz for $H//c$ below $T_{\rm L}$ \cite{Ohta5}. 
As the AFMR appears below $T_{\rm L}$, we can attribute the observed AFMR to ${\rm Cu_A}$ sites. 
The observed results show that the temperature dependence of resonance field obeys the AFMR theory and the linewidth tends to decrease as temperature decreases, which are similarly observed in the case of a typical 3D antiferromagnet \cite{Ohta5}. 
Therefore, we can consider experimentally that an increase in linewidth with a decrease in temperature below  $T_{\rm N}$ in InCu$_{2/3}$V$_{1/3}$O$_3$, as shown in Fig. \ref{fig:fig5}, is a peculiar feature of the $S$=1/2 honeycomb lattice antiferromagnet, which does not originate from the 2D nature because it is not observed in the $S$=1/2 square lattice antiferromagnet Ba$_2$Cu$_3$O$_4$Cl$_2$. 
The suggested studies that can support our claim are as follows: a theoretical study about the spin dynamics in the $S$=1/2 honeycomb lattice antiferromagnet and studies involving the neutron diffraction measurement of InCu$_{2/3}$V$_{1/3}$O$_3$ to investigate the spin dynamics and ordered state below $T_{\rm N}$.

High-frequency, high-field ESR measurements of powder and magnetically aligned samples of the $S$=1/2 honeycomb lattice antiferromagnet InCu$_{2/3}$V$_{1/3}$O$_3$ have been performed in the temperature region from 1.8 to 265 K up to 16 T. 
Although the field of our high-frequency ESR shows the typical behavior of  antiferromagnetic resonance, the linewidth of our high-frequency ESR continues to broaden below $T_{\rm N}$. 
We conclude that the observed anomalous spin dynamics below  $T_{\rm N}$ is the peculiar feature of the honeycomb lattice antiferromagnet, which is not observed in the square lattice antiferromagnet. 

\begin{acknowledgments}
One of authors (S.O.) would like to thank Prof. T. Sakai (JAEA) for the discussion concerning the spin fluctuations of honeycomb and square lattice antiferromagnets. 
This work was partly supported by a Grant-in-Aid for Creative Scientific Research (No. 19654051)  from the Japan Society for the Promotion of Science and Grants-in-Aid for Scientific Research on Priority Areas (No. 17072005 ``High Field Spin Science in 100T'', No. 19052005 ``Novel States of Matter Induced by Frustration'') from the Ministry of Education, Culture, Sports, Science and Technology of Japan.
\end{acknowledgments}


\end{document}